\documentclass[aps,twocolumn,showkeys,showpacs]{revtex4}

\usepackage{graphicx}

\begin{document}

\title{Different topologies for a herding model of opinion}

\author{V.~Schw\"ammle}
 \altaffiliation[Also at ]{Institute for Computational Physics, University of Stuttgart, Pfaffenwaldring 27, 70569 Stuttgart, Germany.}
\author{M. C.~Gonz\'alez}%
 \altaffiliation[Also at ]{Institute for Computational Physics, University of Stuttgart, Pfaffenwaldring 27, 70569 Stuttgart, Germany.}
\author{A. A.~Moreira}%
\author{J. S.~Andrade Jr.}%
\author{H. J.~Herrmann}%
 \altaffiliation[Also at ]{IfB, ETH H\"onggerberg, HIF E12, 8093 Z\"urich, Switzerland.}
\affiliation{%
Departamento de F\'\i sica, Universidade Federal do Cear\'a, 60451-970 Fortaleza, Cear\'a, Brazil.\\
}%


\begin{abstract}
Understanding how new opinions spread through a community or
how consensus emerges in noisy environments can have a
significant impact in our comprehension of the social
relations among individuals. In this work a new model for
the dynamics of opinion formation is introduced. The model
is based on a non-linear interaction between opinion vectors
of agents plus a stochastic variable to account for the
effect of noise in the way the agents communicate. The
presented dynamics is able to generate rich dynamical
patterns of interacting groups or clusters of agents with
the same opinion without a leader or centralized control.
Our results show that by increasing the intensity of noise,
the system goes from consensus to a disordered
state. Depending on the number of competing opinions and the
details of the network of interactions, the system displays
a first or a second order transition. We compare the
behavior of different topologies of interactions: 1d
chains, annealed and complex networks.
\end{abstract}


\maketitle

\section{Introduction}
\label{sec:intro}

An interesting application concerning the structure of
social networks~\cite{Albert2002,Newman2003} is the
modeling of the dynamics of opinion formation. Specific
measurements that cha\-rac\-te\-ri\-ze the statistics behind
the existence of different groups and affiliations within
human populations, justify to model such
aspect of human behavior. The idea behind this field is to
find simple rules of interactions behind the nodes or {\it
agents}, each of which carries its own changing color or
{\it opinion}, trying to reproduce the emergence of complex
patterns observed in reality.

Such opinions can be defined by a finite number of integers
as in the model proposed by Sznajd et al.~\cite{Sznajd2000} or
can even be represented by {\it real} numbers, having a rich
spectrum and opening the possibility for having as many
opinions as agents; like in the model proposed by Deffuant
et al.~\cite{Deffuant2001}. In both cases the proposed dynamics
has a natural absorbing state or {\it consensus}, in which
all the agents share the same opinion.
Other models as the one by Hegselmann and Krause~\cite{Hegselmann2002},
the voter model~\cite{Krapivsky2003}, the Galam's majority
rule~\cite{Galam2000}, and the Axelrod's model~\cite{Axelrod97}
are reviewed in Ref.~\cite{Boccaletti2006}. As pointed out there,
very few and non conclusive results exist for the consensus
models on complex networks.

In this work, we present a general model, where o\-pi\-nions
are represented by vectors with real components and the
agents interact with a non-linear rule.
We propose, for the abstract space of human opinions a
dynamical rule where each agent has a opinion vector that is
fixed in modulus. Every time step an agent interacts with its
neighbors and assumes a new value for its opinion
vector that is a function of the average direction of its
neighboring agents plus an added noisy term. The resulting
behavior presents two important characteristics, ({\it{i}})
although the model allows a continuous change from one
opinion to some other, the interaction favors extremely
decided states over undecided states, and ({\it{ii}}) the
system ubiquitously evolves to coordination and grouping
without the need of leaders or centralized control.
The fact that the modulus of the opinion vector is constant
describes the strength of an opinion about a
specific topic at the expense of the other beliefs.  According
to our model, undecided agents (i.e., those that do not have
a strong belief in one particular opinion) can not affect the
ones with a strong opinion. This type of interaction is
somewhat different to the one used in models of opinion
formation, which usually consists in weighted
averages. Similar rules have been proposed for example in
Ref.~\cite{Baronchelli2006} to explain how very large
populations are able to converge to the use of a particular
word or gram\-ma\-tical construction without global
coordination.

We study the transition to consensus as a function
of noise. As a new feature, we find that
different types of transitions with or without hysteresis
are observed depending on the dimension of the opinion vector.
Additionally, we observe that the transition is controlled by the interaction dynamics 
and is independent on the correlations of the network topology.

We start in Section~\ref{sec:orig} by describing in detail the
proposed model for interactions between opinions. In
Section~\ref{sec:res}, we present the results of the
transition to consensus as a function of the noise for 1d
chains, annealed, and complex networks. Conclusions are
given in Section~\ref{sec:concl}.

\begin{figure}[!h]
  \begin{center}
    \includegraphics[angle=270,width=0.8\columnwidth]{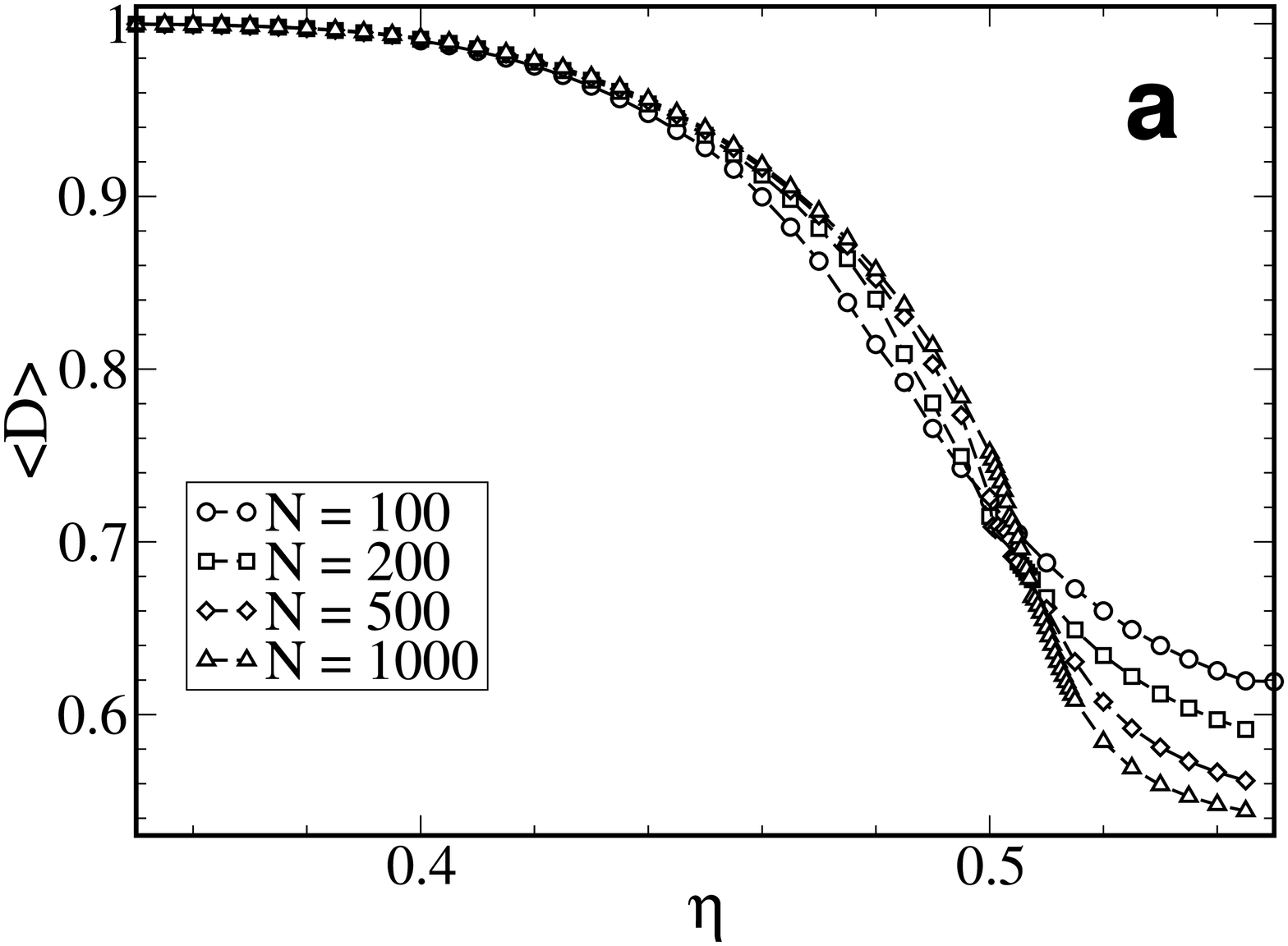}
    \includegraphics[angle=270,width=0.8\columnwidth]{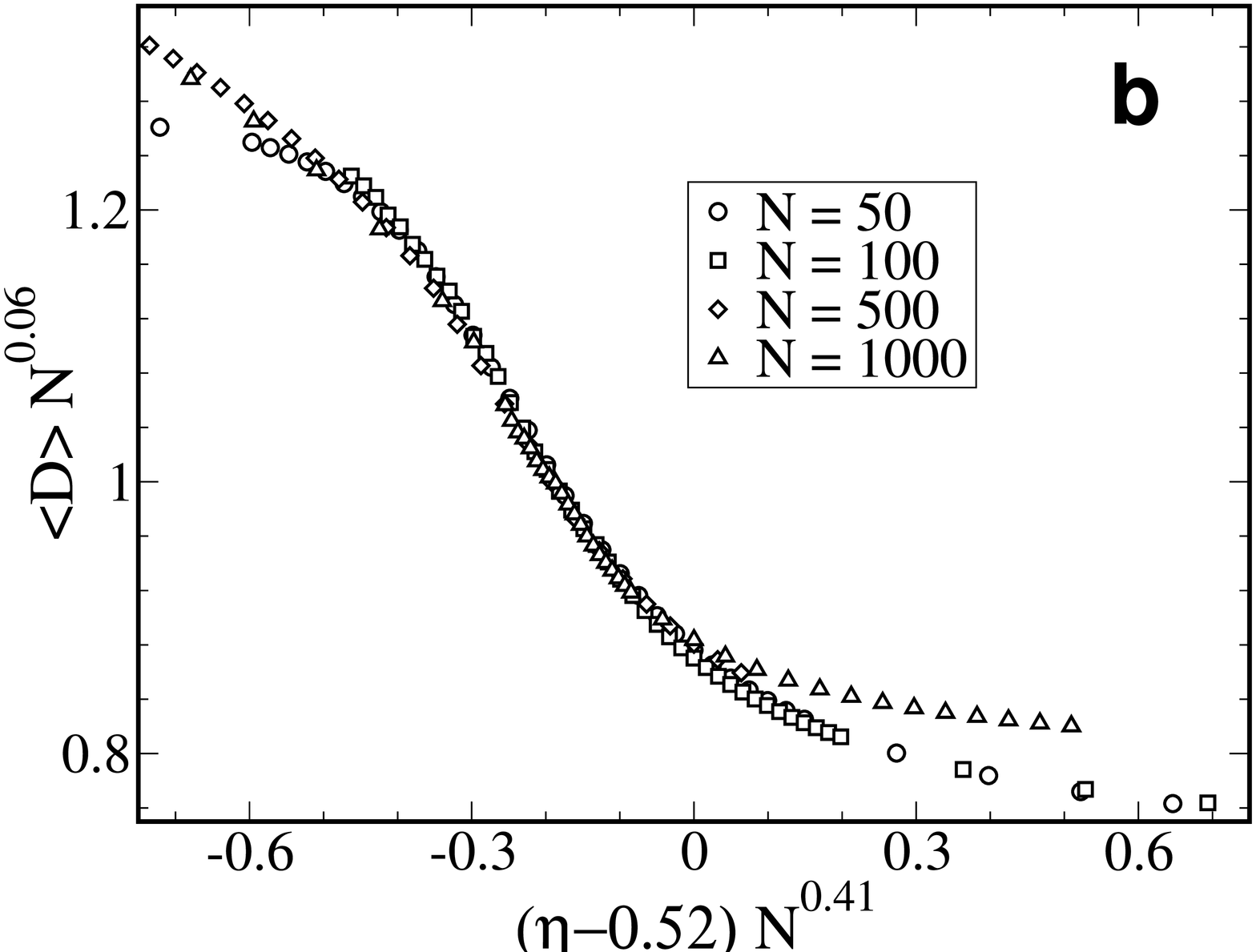}
\caption{Finite size scaling of the transition in a system with a two-dimensional opinion vector with random
interactions. The figure compares the frequency of agents with the same dominant opinion versus
the noise $\eta$ for different population sizes $N$. {\bf a}: Original data. {\bf b}: Finite size
scaling. Each point corresponds to an average over $10-20$ runs with different random seeds.}  
    \label{fig:MeanScaling}
\end{center}
\end{figure}

\section{The opinion model}
\label{sec:orig} 

The system comprises a fixed number of $N$ agents.
Every agent $i$ is characterized by its own opinion vector $a_n^{(i)}$ of $n=1,..,O$ opinions. Each
element of this vector corresponds to a different opinion about the same topic. Negative
  values of the elements are not allowed. The sum of the vector elements of an agent is one:
    $\sum_{n=1}^O a_n^{(i)}=1$. For instance, agent
$j$ favors communism with 20\% and capitalism with 80\% (given $O=2$): $a_1^{(j)}=0.2, 
a_2^{(j)}=0.8$. Each time step every agent actualizes its opinion vector by comparing its
values to the ones of its $k_i$ nearest neighbors.
These other agents are chosen by the topology
of the graph, and the agent updates its opinion vector due to the following rule,
\begin{equation}
\hat{a}_n^{(i)}(t) = \sum \limits_{l=1}^{k_i} a_n^{(i)}(t) a_n^{(l)}(t) + k_i g(t),
\label{eq1}
\end{equation}
where $g(t)$ presents a stochastic variable, distributed uniformly in the interval
$[0,\eta]$. With this exclusively positive noise we assure that $a_n^{(i)}(t)\geq 0$.
This stochasticity can be interpreted to be due to misunderstandings among the 
agents, the spread of wrong information, or other perturbing actions.

The interaction term in this model is of second order. Thus, in a noiseless
environment, the agents tend to have the same stronger opinion.
The factor $k_i$
avoids that agents which have more connections feel less noise. 
In order to guarantee that the sum of opinions is equal to one, 
the vector is normalized afterwards similar to the model presented in Ref.~\cite{Lorenz2006},
\begin{equation}
a_n^{(i)}(t+1) = \frac{\hat{a}_n^{(i)}(t)}{\sum \limits_{m=1}^{O} \hat{a}_m^{(i)}(t)}.
\label{eq2}
\end{equation}
In order to elucidate the principal properties of the
update rule given by Eqs.~(\ref{eq1}) and ~(\ref{eq2}), we
examine in detail the noiseless interactions between three
types of agents with different characteristic values of a
two--dimensional opinion vector ($O=2$), namely,
$a_1=\{0.8,0.2\}$, $a_2=\{0.5,0.5\}$ and
$a_3=\{0.2,0.8\}$. First, an interaction between an agent having $a_1$ with another having
  the same $a_1$ results in
$\{0.94,0.06\}$---interactions between agents with the same
dominant opinion strengthens their belief in this opinion.
$a_1$ with $a_2$ yields $\{0.8,0.2\}$---interactions with
``undecided'' agents are ineffective in the sense that
agents without dominant opinion are not able to convince
another agent. On the other hand, this interaction will have
a substantial effect on the undecided agent, i.e., undecided
agents are convinced easily.  The interaction between $a_1$
and $a_3$ results in $\{0.5,0.5\}$---interactions of agents
with opposite opinions lead them to become less decided.

At the beginning of a simulation the opinion vectors are initialized either randomly or by
consensus: for random initialization we toss for each opinion component of each agent a number
between zero and one. The opinion vectors are normalized afterwards according to Eq.~(\ref{eq2}).
The other way to initialize the system (consensus) is by setting one to the first element of
each vector and fill the rest with zeroes. 

The main parameter of this model is given by the maximal noise $\eta$ which we will call from
now on the control parameter. Its role corresponds to the one of a temperature in physical
systems. In a social system, the noise represents any internal or external interference in the
communication among the agents. Other free parameters of the system are given by the number of
agents $N$, the number 
of opinions $O$ and the number of agents $k_i$ to interact with per time step. The last parameter
can be different for distinct agents depending on the topology of the actual network.

\begin{figure}[!h]
  \begin{center}
    \includegraphics[angle=270,width=0.9\columnwidth]{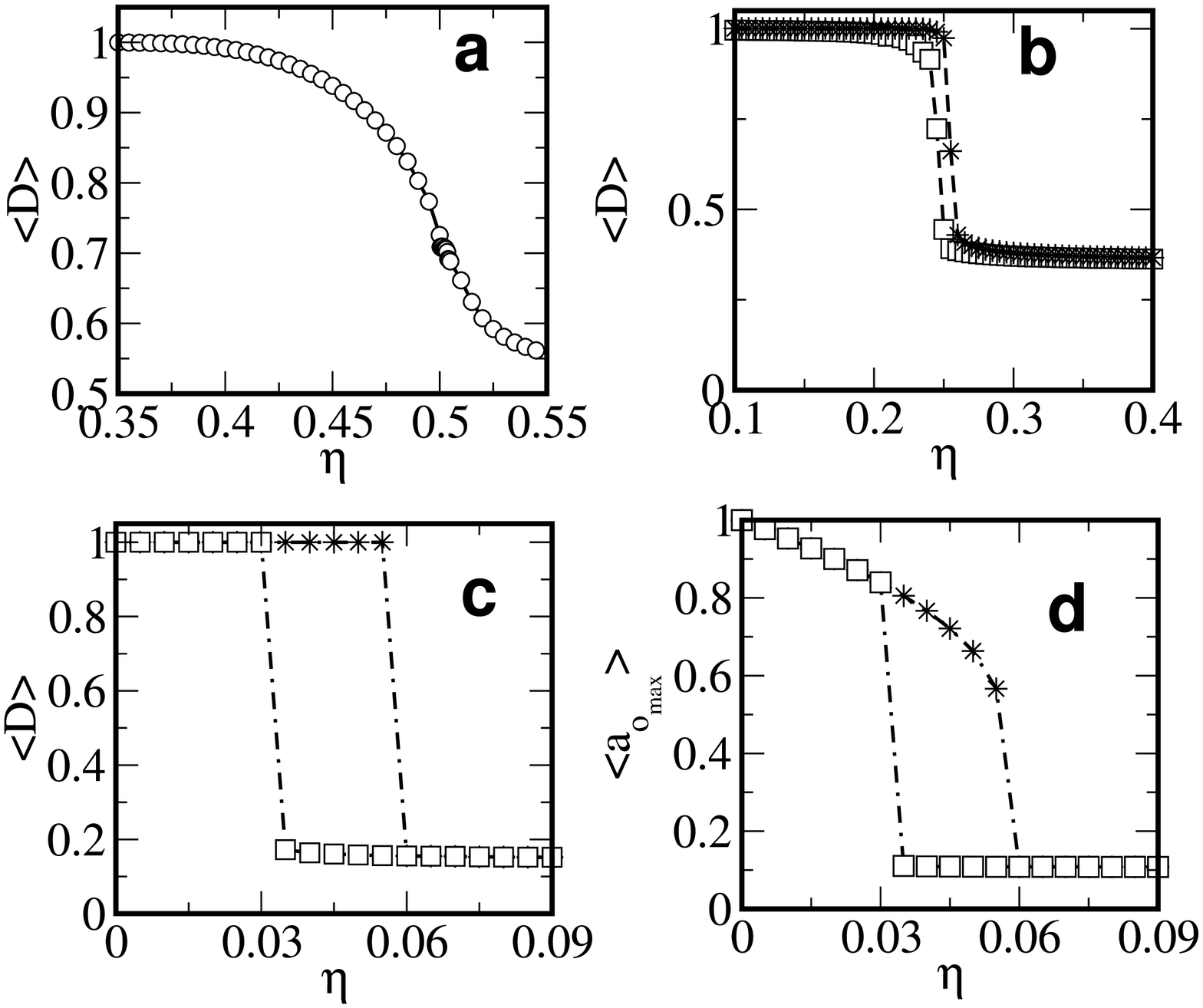}
\caption{The transition of the system for different numbers of opinions $O$ and
different initial conditions. The population size is 500. {\bf a}: The outcome of a system of two
opinions (circles) is independent of its initialization. {\bf b}: In the case of three opinions
(dashed line), the curves present a hysteresis and the results are different, if the field
is initialized randomly (squares) or with the system being already in the consensus state
(stars). {\bf c}: A system with an opinion vector containing ten opinions also exhibits hysteresis
(dash-dotted line).{\bf d} Average value of the dominant
opinion $a_{o_{max}}$ vs. $\eta$ for the same simulation as in {\bf c}.} 
    \label{fig:MeanHyst}
\end{center}
\end{figure}
%


A simple mean-field solution of the model without noise can be derived. Suppose a state where all agents have the same
values in their opinion vectors. Thus the index of the agents can be suppressed, $a_n^{(i)}(t)
= a_n(t)$ and $a_n^{(i)}(t+1) = a_n(t+1)$, and Eqs.~(\ref{eq1}) and~(\ref{eq2}) can be summarized. In
the case of two opinions the equations correspond to the map,
\begin{equation}
a_1(t+1) = \frac{a_1^2(t)}{a_1^2(t)+a_2^2(t)}\, , \,\,\, a_2(t+1) = \frac{a_2^2(t)}{a_1^2(t)+a_2^2(t)}.
\end{equation}
The fixed points of these equations are $(a_1,a_2) \in \{(0,1),(1,0),(0.5,0.5)\}$ where the first two
ones are stable and the last one is unstable.
The solutions for $O$ opinions are in $\{(1,0,0,...),(0,1,0,0,...),...,(0,0,...,0,1)\}$ with all
$a_n$ stable. All other solutions have at least one unstable element of the opinion vector and
thus the unstable element influences the other ones until an absorbing state
with one opinion totally dominant is reached. 
 
\section{Results}
\label{sec:res}

\subsection{Annealed interactions}
First, we present simulations of the model without fixed topology. Each time step, a
simulation runs over all agents. For each of them and at each time step, two new
random partners are chosen to interact. 
We chose interaction with two other agents ($k_i=k=2$) in order to
facilitate the comparison of this case with the one of a one-dimensional chain which will be
explained in the following section. The
annealed approach avoids long term behavior and the distribution of opinions
reaches the stationary state fast. Because the interacting units are a sampling
of the whole system, it is expected that this annealed approximation
should behave similar to a mean-field.

\begin{figure}[!h]
  \begin{center}
    \includegraphics[angle=270,width=0.9\columnwidth]{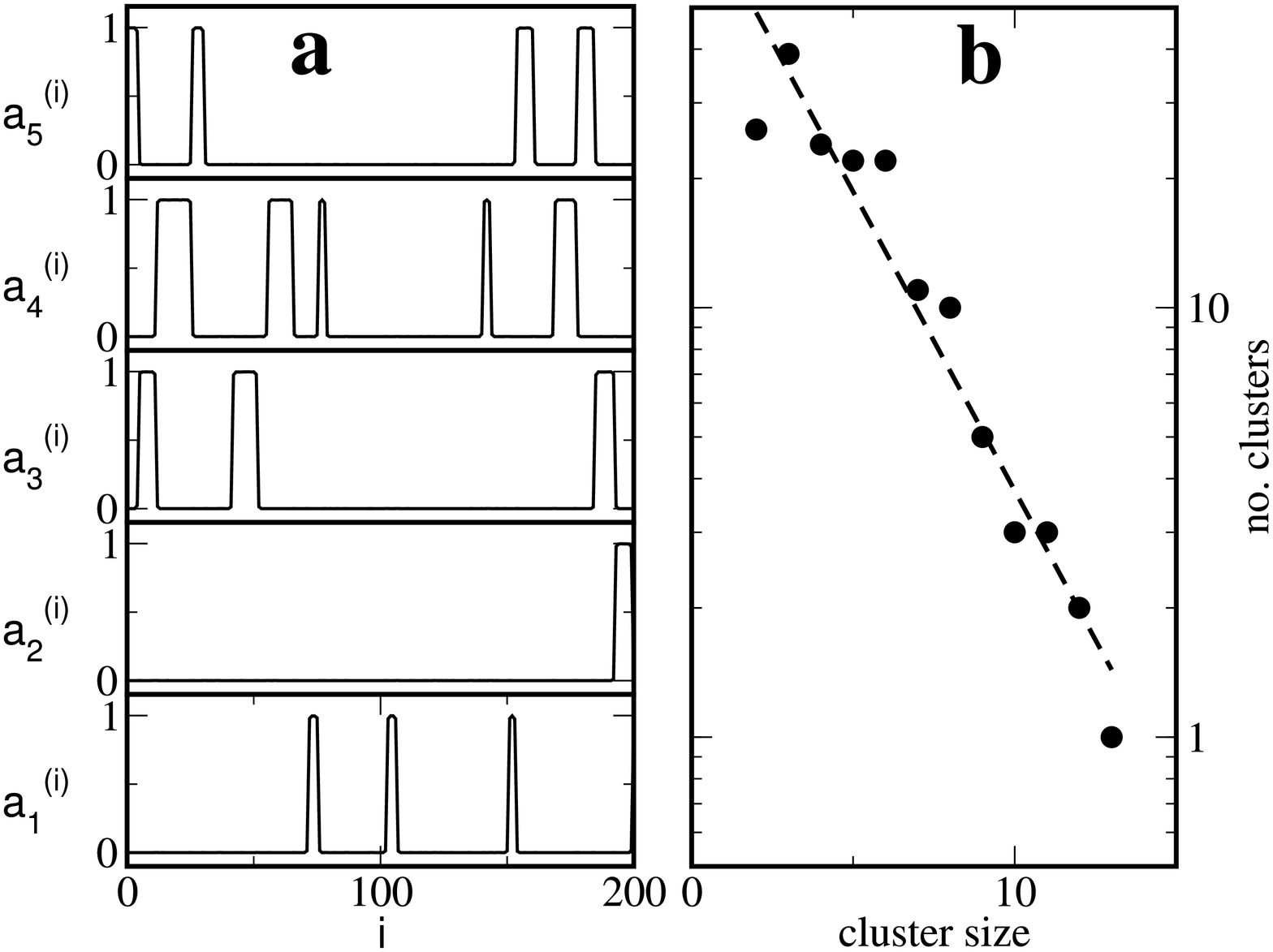}
\caption{Ten opinions on a one-dimensional chain. These results illustrate simulations
without noise ($\eta=0$) and a system of $1,000$ agents. {\bf a}: The figure zooms on the first
200 agents of the population, where each line corresponds to a different element of the
opinion vectors. Only the first five opinions are displayed. The agents form local clusters of
different dominant opinions. {\bf b}: The cluster sizes distribute following an exponential decay. }
    \label{fig:Ring10Op}
\end{center}
\end{figure}

The results reveal that the system can reach two different absorbing states. At small values of
the control parameter (maximum noise 
$\eta$), one opinion completely dominates the system, $o_{max}$. 
For a noise $\eta$ larger than a certain value, each opinion remains 
with the same frequency, $1/O$. The order parameter $D$ is 
the frequency of the agents which have an opinion vector with 
the same dominant opinion, being itself dominant in the
system. More precisely: for each agent we search its strongest opinion and
then count, for each opinion, the number ``$n$'' of agents with this opinion
as their dominant one. The largest value $n_{D}$, and so the most dominant
one of the system, determines $D=n_{D}/N$. $\langle D \rangle$ means, that we
average $D$ over many time steps. This order parameter is
normalized, so that it is unity if all agents have 
the same dominant opinion, a state we call the consensus state. The value $1/O$ corresponds
to a uniform distribution of opinions. A transition occurs between 
consensus and uniform distribution, when $\langle D \rangle$ goes from $1$ to
$1/2$ in the case of two opinions (Fig.~\ref{fig:MeanScaling}a).

\begin{figure}[!h]
  \begin{center}
    \includegraphics[angle=270,width=0.9\columnwidth]{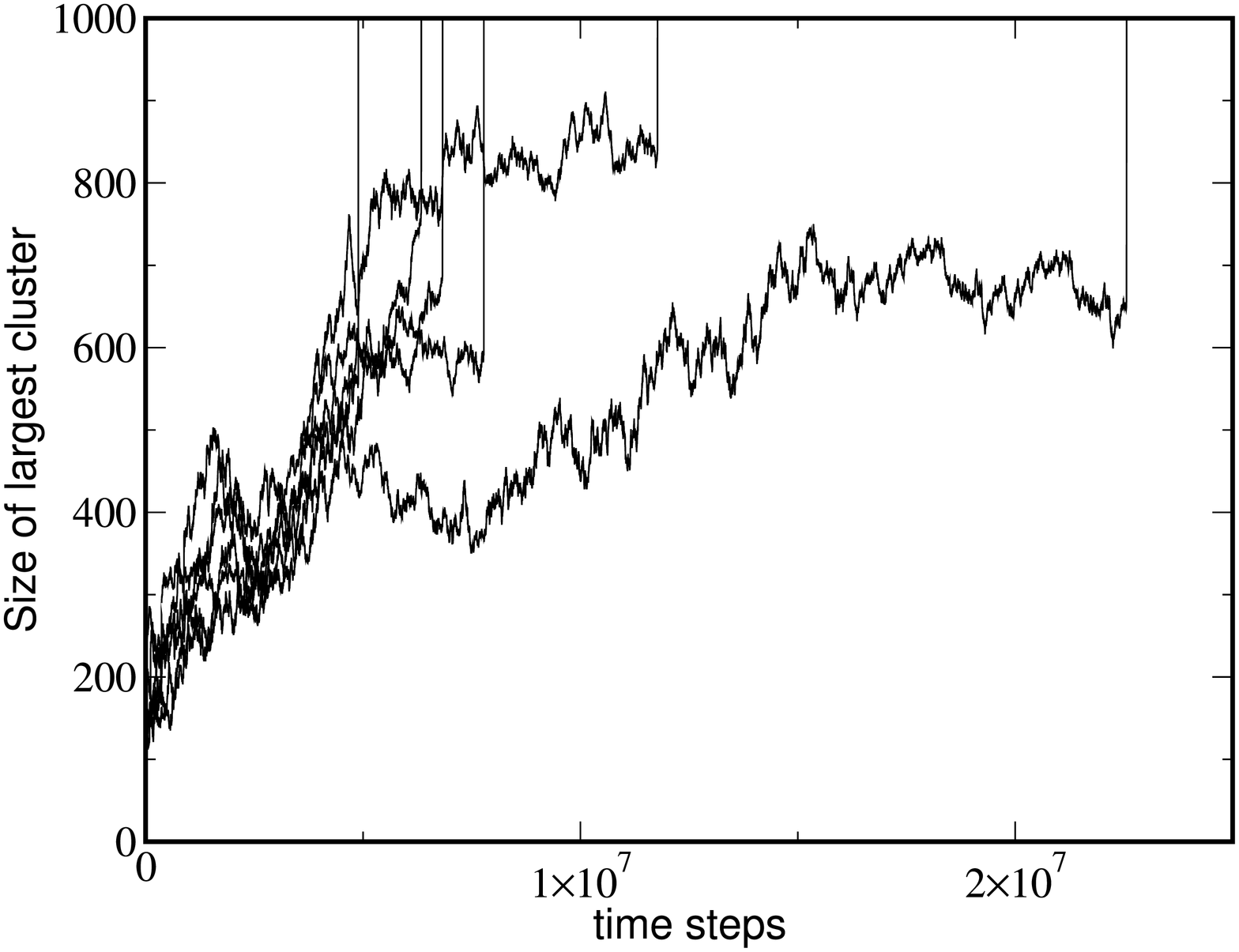}
\caption{The size of the largest cluster increases in time until it
reaches the population size $N=1,000$. Here, we see the results of different random initializations of the system
on a chain. The number of opinions is ten, and the noise $\eta=0.2$.}
\label{fig:MaxClusterRing}
\end{center}
\end{figure}

The transition becomes more abrupt for larger population
sizes. A transition point characteristic for the jump 
from the consensus to the uniform
states is located at $\eta_c \approx 0.5$, increasing with the population size. 
This transition seems to be a phase transition of
second order. We carried out finite size scaling in order to
obtain the critical exponents (Fig.~\ref{fig:MeanScaling}b). 
Near the critical point the curves coincide using the scaling relations,
\begin{equation}
 \langle D \rangle N^{\frac{-\beta}{\nu}} = (\eta - \eta_{c}) N^{\frac{1}{\nu}},
\label{fin1}
\end{equation}
with $\nu \approx 2.4 \pm 0.1$, $\beta = 0.15 \pm 0.05$, and the
critical noise $\eta_c = 0.52$. 

\begin{figure*}[!t]
  \begin{center}
    \includegraphics[angle=0,width=0.9\textwidth]{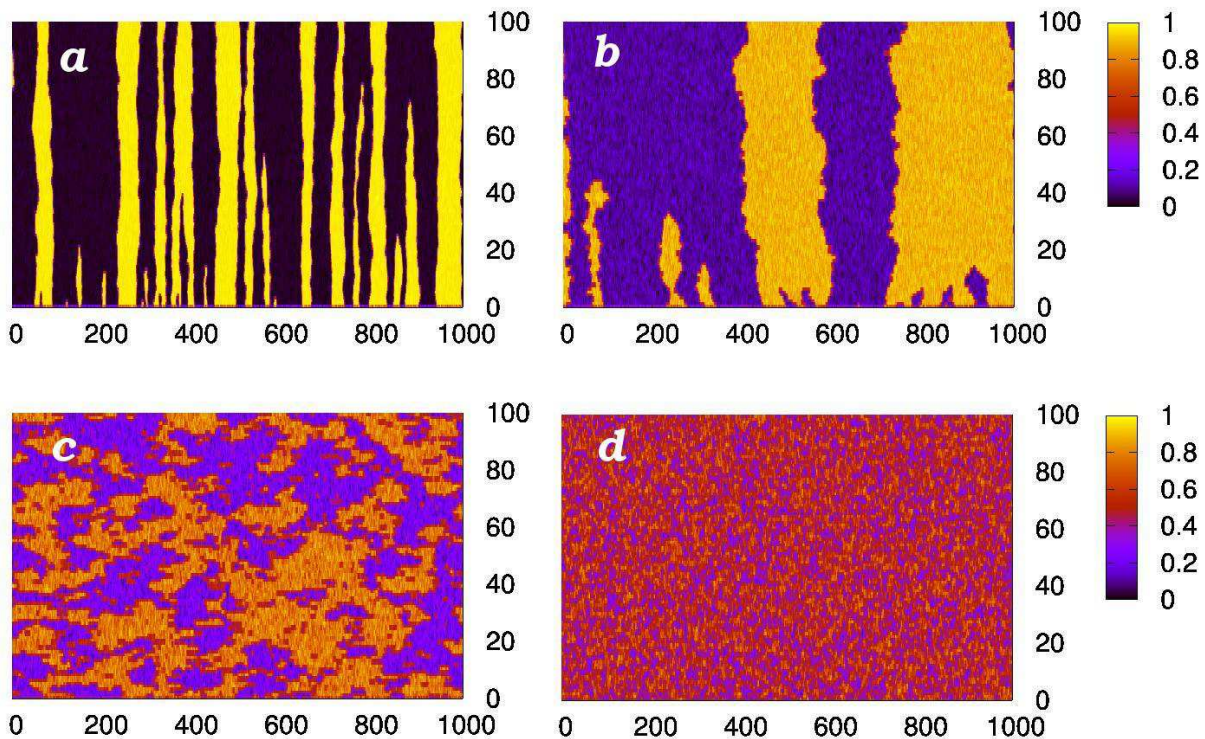}
\caption{(Color online) Graphical illustration of the temporal behavior of the system on a chain. The color
(gray tone) 
corresponds to the value of the first of the two opinions of the system. The simulations run
over $100,000$ time steps, drawing each $1,000$ iterations a new point on the vertical axis
beginning at the  bottom. The horizontal axis depicts the location of each agent on the chain,
altogether consisting of $1,000$ agents. The noise $\eta$ is 0.05 in (a), 0.2 in (b), 0.35 in
(c), and 0.45 in (d). }
    \label{fig:ImageRing}
\end{center}
\end{figure*}

Fig.~\ref{fig:MeanHyst} shows that in the case of annealed interactions the
transition becomes of first order for simulations with an
opinion vector of more than two opinion elements. 
The fluctuations do not increase at the transition point. Now, the
transition from the consensus to the uniform state depends on the 
initialization and is much more abrupt. If the initialization is random, the
system jumps to the consensus state at lower values of $\eta$ than in 
the case of an initialization with consensus in one opinion. A transition
with a typical hysteresis occurs at lower values of $\eta$ if we increase the
dimension of the opinion vector. 

Note that $D$ gives us the fraction of agents with dominant opinion ($o_{max}$)
but does not contain information about $a_{o_{max}}$, 
the magnitude of the component associated to $o_{max}$. 
In Fig.~\ref{fig:MeanHyst}d we plot $a_{o_{max}}$ vs. $\eta$
for the same simulations presented in Fig.~\ref{fig:MeanHyst}c.
$a_{o_{max}}$ is larger for lower values of $\eta$ and 
below a certain $\eta_{c}$ consensus is observed for both kinds
of initializations, {\it only} when a large value of $a_{o_{max}}$ is
reached. 
This is a nice feature of our model: {\it consensus and
resolution emerge together in the system}. That is, the
agents can only make up their minds for a preferred opinion when
consensus is achieved through all the system.

\subsection{One-dimensional topology}

If we put the agents on an one-dimensional lattice with periodic boundary conditions, or, in
other words, a chain, the results become different. First, we concentrate on the case of ten
opinions and no noise ($\eta=0$): The system is now highly 
dependent on the initial state. A random initialization of the opinion vectors leads to the
situation depicted in Fig.~\ref{fig:Ring10Op}a. The same amount of each opinion seems present in
the system during the evolution. The system organizes itself by rearranging its opinion vectors to form local
clusters of different sizes. In one cluster the same opinion dominates for all
agents. Each agent has a well pronounced dominance of an
opinion (its value being nearly one), and the interfaces between clusters of different dominant
opinions are very sharp. These clusters develop fast after the
beginning of the simulation. The distribution of cluster sizes follows the exponential decay of
a Poisson distribution (Fig.~\ref{fig:Ring10Op}b). The results
with $\eta=0$ are qualitatively the same for different numbers of opinions, $O$.

\begin{figure}[!h]
  \begin{center}
    \includegraphics[angle=270,width=0.95\columnwidth]{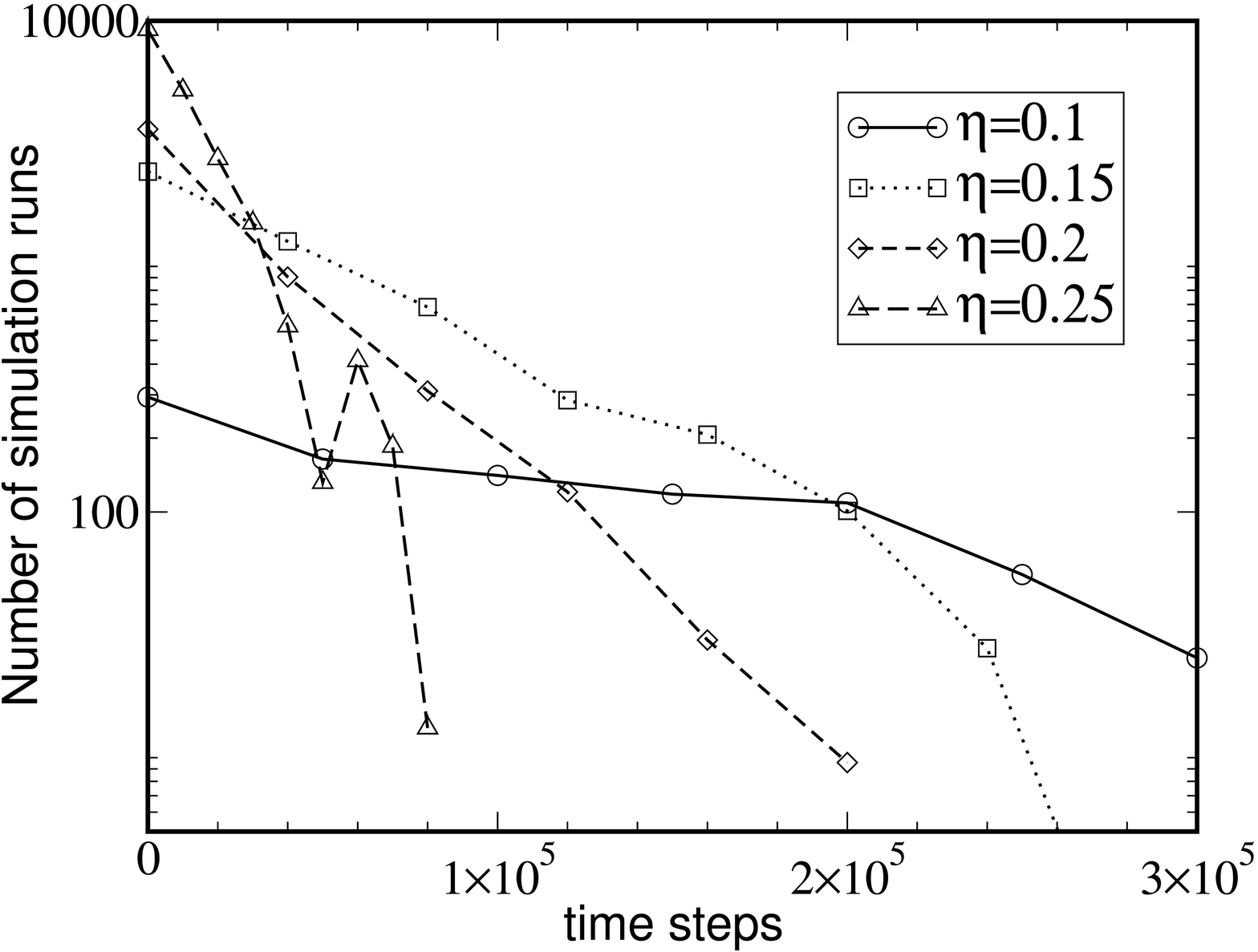}
\caption{Histogram of the time steps needed to reach the consensus state. Each curve corresponds to
simulations with the same parameters: 100 agents, 2 opinions, random initialization. For
each value of the noise parameter we carried out $200,000$ runs with different random seeds.}
    \label{fig:TimeHisto}
\end{center}
\end{figure}

Noise ($\eta>0$) leads to a slow increase of one of the ten opinions
with time. The dominant opinion absorbs more and more of the losing
opinions. Fig.~\ref{fig:MaxClusterRing} 
illustrates how the largest cluster of the system temporally evolves for $\eta=0.2$. 
As also can be recognized in this figure, the
time to reach consensus can be really long, even in a 
small system of $1,000$ agents.

\begin{figure}[!h]
  \begin{center}
    \includegraphics[angle=270,width=0.9\columnwidth]{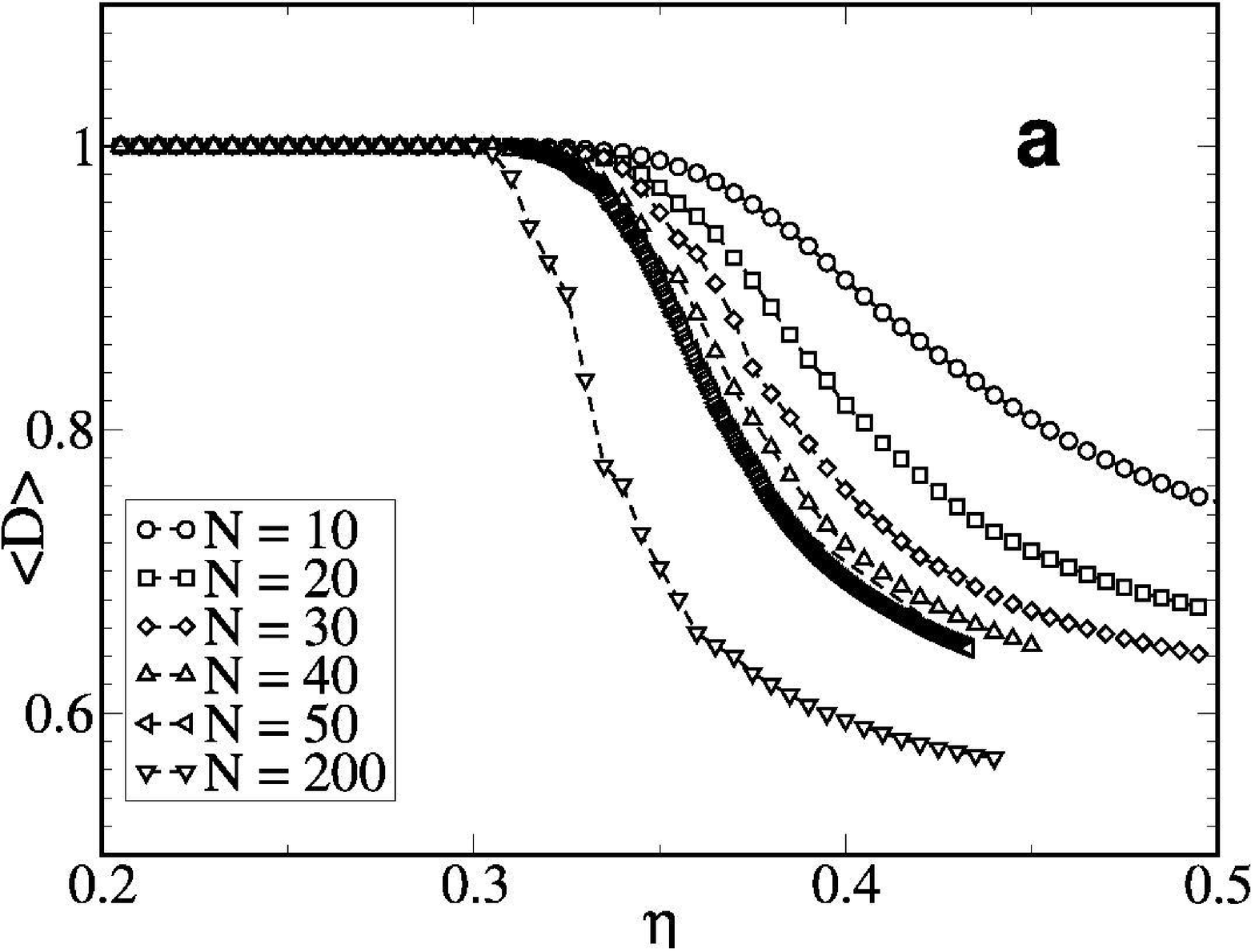}
    \includegraphics[angle=270,width=0.9\columnwidth]{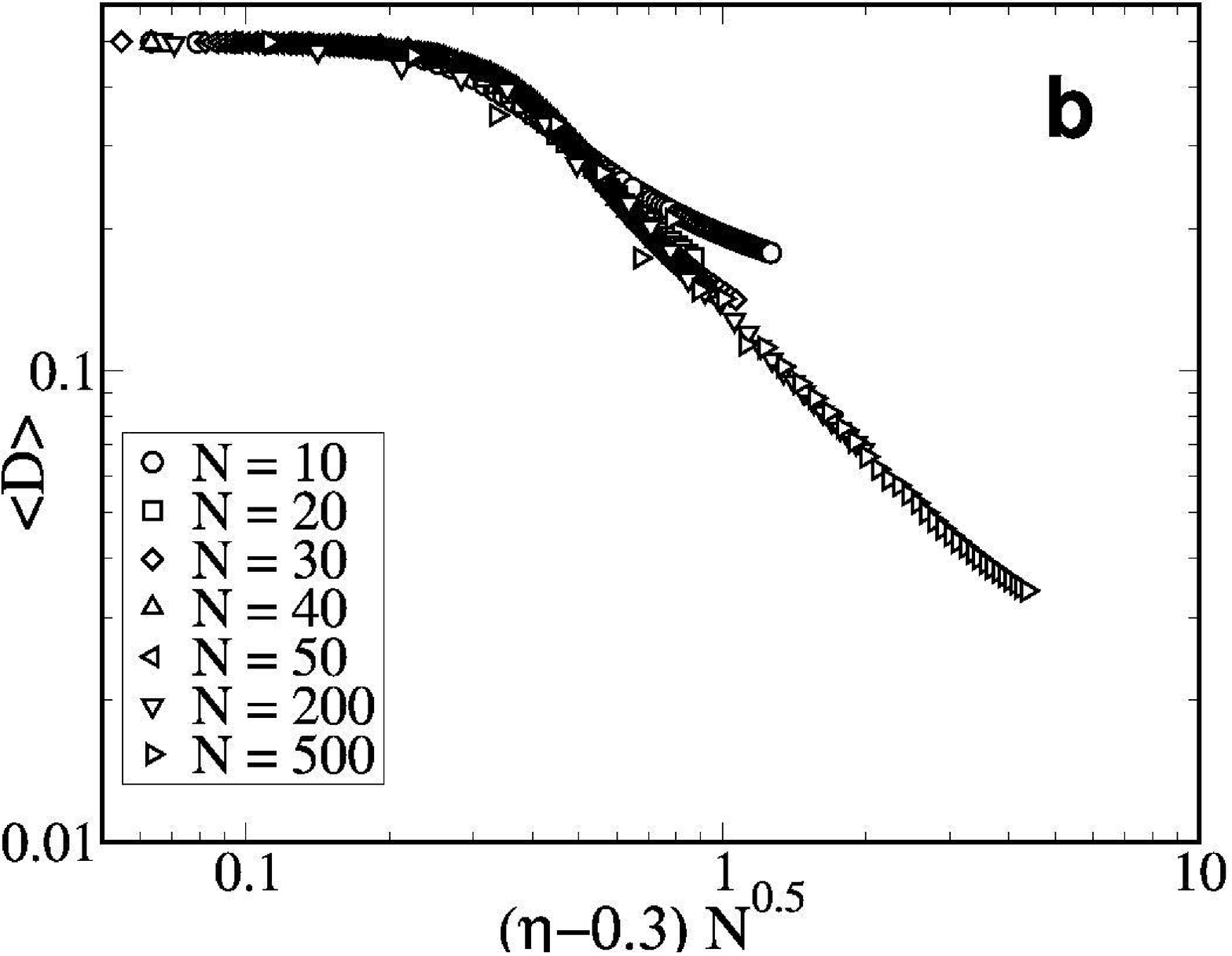}  
\caption{{\bf a}: The fraction of agents with the same dominant opinion versus
$\eta$ are compared for different population sizes, $N$, in a system of two opinions, where
only the nearest neighbors on a chain interact. {\bf b}: Finite size scaling of the transition.}
    \label{fig:RingNMean}
\end{center}
\end{figure}

With non-zero but small noise, the information propagates slowly through the
sample. Because of that, the time to reach the absorbing state is much larger than in the
case of random interactions. Next we consider a system consisting
of $1,000$ agents, which have opinion vectors of two
dimensions. The normalization of Eq.~(\ref{eq2}) allows us to focus only on the temporal
behavior of one of each agent's opinion without loss of
information. Fig.~\ref{fig:ImageRing}  exhibits this time behavior for the noises
$\eta=0.05,0.2,0.35,0.45$ during the first $100,000$ time steps. Each agent's first opinion
is depicted by a color (gray tone)  which corresponds to its value, and evolves beginning at the bottom. 
At low noise values stable clusters seem to form. The size of the clusters becomes smaller with decreasing
$\eta$. Nevertheless, these clusters are not stable, and the system reaches the consensus state
after a finite time. For $\eta=0.05$ and $\eta=0.2$ the size of clusters with the second dominant opinion is
larger, indicating that at the end this opinion will control the system. The larger the size of
a cluster the longer it takes to break it. At values of $\eta$ larger than $\eta=0.3$, strong
fluctuations control the system, and consensus begins to become unstable. 
For values around $\eta = 0.3$ one opinion still dominates, and clusters appear and
disappear. At larger $\eta$ the opinions have values around $0.5$ for all agents which do
not fluctuate much.

It is interesting to calculate the number of time steps the system needs to reach its final
state.  In a system of two opinions we carried out various simulations with the same value of noise, $\eta$,
and a population size of $N=100$. Each simulation begins with an initialization of randomly
distributed opinions but a different random seed. Fig.~\ref{fig:TimeHisto} shows the 
distribution of times, needed to reach the consensus state for a system. The distribution decreases 
exponentially. The distribution becomes broader with decreasing values of
$\eta$, where $\eta=0$ should correspond to a flat distribution. 

As in the case of random interactions, a transition occurs from the consensus
state to the one of a uniform distribution of the opinions.
Fig.~\ref{fig:RingNMean}a 
illustrates the variation of $\langle D \rangle$ with $\eta$ for different
system sizes. As shown in Fig~\ref{fig:RingNMean}b, by performing a finite-size scaling
analysis through Eq.~(\ref{fin1}), the collapse of all curves is obtained when we use the critical
exponents  $\nu = 2$, and $\beta = 2$.

\begin{figure}[!h]
  \begin{center}
    \includegraphics[width=0.9\columnwidth]{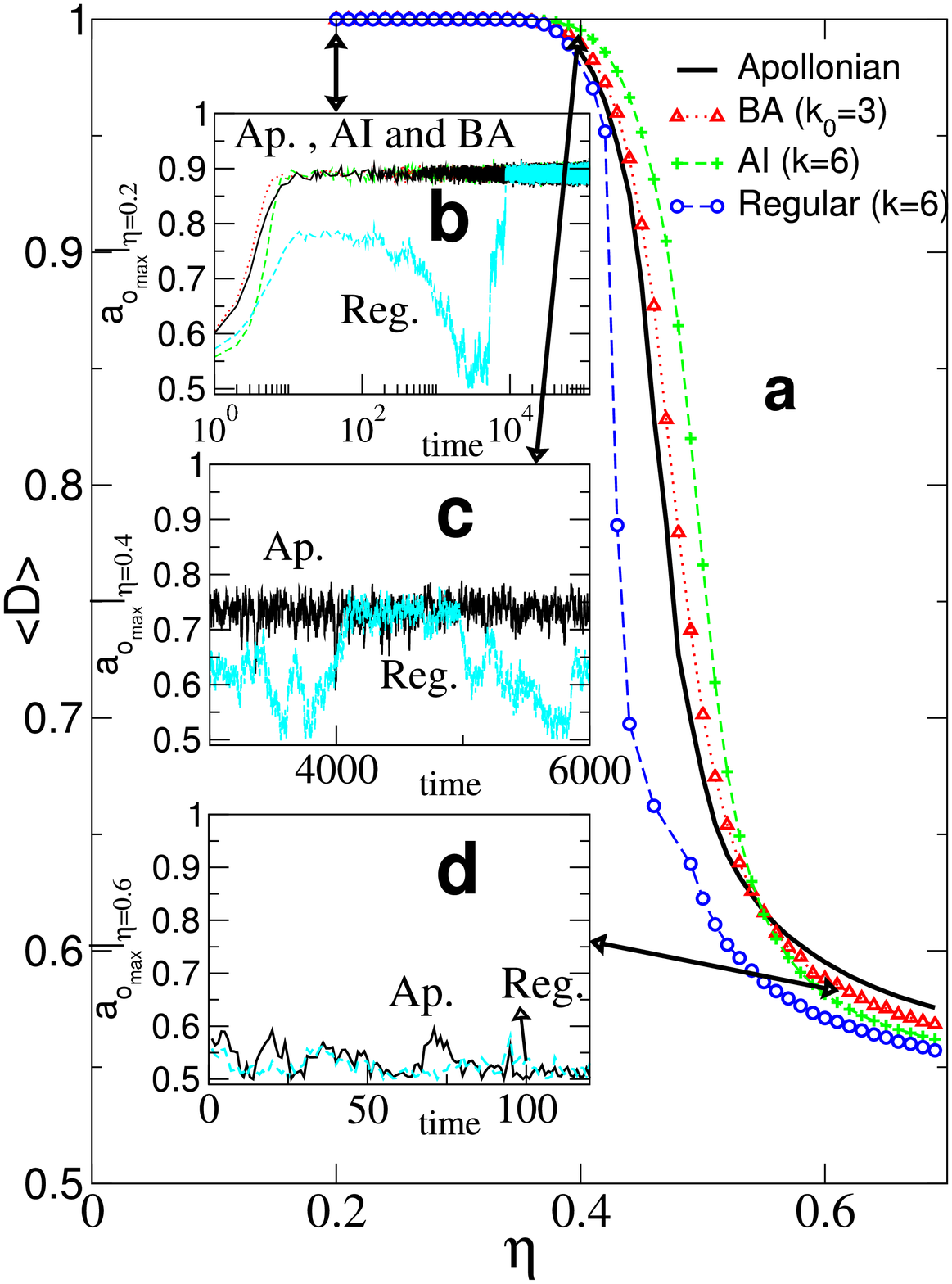}
\caption{(Color online) {\bf a:}Influence of the topology of networks 
on the transition to consensus ($D=1$) as a function of noise ($\eta$).
The transition on two different scale-free networks, the Apollonian (solid
line) and the Barab\'asi-Albert network (BA, triangles) is similar to the one
observed for annealed interactions (AI, plus signs), and differs from the 
transition on a regular lattice (circles). In the three insets we plot 
the value of the dominant opinion, $a_{o_{max}}$ vs. time. 
{\bf b:} Comparison
of the behavior of $a_{o_{max}}(t)$ on the four networks: 
Apollonian (solid line), Barab\'asi-Albert (BA, dotted line), Annealed
(AI, dashed-line) and Regular (long dashed line) for a fixed noise ($\eta=0.2$).
One observes that for this noise, which is below the critical
noise, in the regular network the emergence of consensus takes 
longer than in scale-free and annealed interactions, which have similar 
behavior (three upper curves). 
{\bf c:} Near below
the transition, for $\eta=0.4$, we compare the response of the regular 
and the Apollonian network. It is observed that for the former there is
an intermittency among consensus $D=1$ and $a_{o_{max}}=0.79$  
and not consensus $D=0.5$ and $a_{o_{max}}=0.5$. This behavior is not 
observed in the complex networks. 
{\bf d:} Above 
the transition ($\eta=0.6$) the consensus is broken and the dynamics of the 
opinion $a_{o_{max}}$ vs. time behaves similar in regular and complex
networks. All simulation runs are with systems of 124 agents.}  
\label{fig:sf}\end{center}
\end{figure}

\subsection{Complex networks}
In this section we compare the behavior of the opinion model 
if the agents interact  with their $k$ nearest neighbors
on different networks topologies. We study two different 
kinds of scale-free networks; i.e. networks with a power
law degree distribution $k^{-\alpha}$. 
Those are the  Barab\'asi-Albert network ($BA$)\cite{Albert2002} and the 
Apollonian network~\cite{Andrade2005}. The networks
have considerable topological  differences, that can be expressed 
in terms of their clustering coefficient $C$. This coefficient is the average probability
that the neighbors of a  node are connected among them.  
The $BA$ network has a clustering coefficient, $C$, which depends
on the network size as $N^{-1}$. It is independent on the degree
of the nodes. In contrast, the Apollonian network has hierarchical
structure with $C$ depending on the degree of the node as a power law of
the degree and its average value is high ($C\approx 0.8$) and independent of the 
network size $N$. Both types of scale-free networks, with and 
without hierarchical structure, have shown to be good models 
for rather different kinds of social interaction networks, 
from social collaboration networks~\cite{Ravasz2003} 
to networks of sexual contacts\cite{Gonzalez2006}. 

Further, we show that despite of the structural differences of these networks,
the formation of consensus depends mainly on the noise and is independent on
the specific topology of the scale-free network studied in the case of two opinions. 
The transition to consensus as a function of noise for 
the two scale-free networks, seems to belong to the same 
type of transition as in the case of annealed interactions. In contrast,  
we compare the behavior of the model with a regular network with
$k=6$ on a chain (in the previous section we had $k=2$), 
adding interactions up to the third nearest neighbors.
The transition from consensus to a uniform distribution on the
regular network 
differs from the transition of complex networks and annealed case and
presents similar behavior as the one reported in previous sections
for a chain.

In Fig.~\ref{fig:sf}a we show $\langle D \rangle$ vs. $\eta$ for the model
on the $BA$ (triangles) and Apollonian networks (solid line),
compared to the result of annealed interactions (plus signs) and the
regular network (circles). The results of the figure represent 
the average over $20$  realizations on systems of $N=124$ 
agents and $2$ opinions. 
Near the transition, the fluctuations on the regular lattice strongly increase,
as opposed to annealed interaction and to $BA$ and Apollonian networks. This is because the system presents an
intermittency near the transition point ($\eta \approx 0.4$).
We observe this intermittency of the dynamics in
Fig.~\ref{fig:sf}c, comparing the value of 
the dominant opinion $a_{o_{max}}$ vs. $time$, for
the Apollonian and the regular network with $\eta=0.4$.
Above the critical noise, there is no consensus and
the fraction of agents with dominant opinion is $\langle D \rangle \approx 1/2$.
At these values of $\eta$, the response of the system is
similar for scale-free and regular networks, 
as is shown in Fig.~\ref{fig:sf}d 
with $\eta=0.6$. 

Above the critical noise there is no way for the agents to
achieve global coordination. In this situation, the dynamics
is dominated by local interactions, thus the topology of the
system has little effect on this regime. Below the critical
noise, global coordination becomes possible. However, the low
dimensionality of the regular lattice leads to the
intermittent behavior observed in the panel~\ref{fig:sf}c.

\section{Conclusion}
\label{sec:concl}

Starting from a model based on interactions with a term of second order, we analyzed its
behavior for different topologies: random, regular and complex ones. Depending on the control
parameter, the noise $\eta$, two different absorbing states control the system. Its behavior
changes from consensus to a uniform distribution of opinions. Despite the rather simple
approach to take into account such simple interactions, a rich variety of results can be
reported depending on the dimension of the opinion vector. 
The results show that an opinion is kept (for systems with more than
two opinions), and the parameters need to be adjusted crucially to change the state (hysteresis).
This occurs at different dimensions $O$ of the opinion vector, depending on the topology of interactions.

The response of the system to approach consensus has
the origin in the model dynamics as opposed to the particular
features of the network. An important
characteristic of the transition to consensus is the dimension
associated to the space of agent interactions.
The dynamical response of the opinion model for both scale-free networks
is similar to the response observed for annealed interactions and each of
these cases represents long range interactions. In contrast,
differences are reported with a regular lattice, which has spatial
dimension one, associated to nearest neighbors interactions.

As it was previously observed for the Sznajd model
of opinion formation,
for the general model that we
present here, the response of the system in terms of
opinion formation is qualitatively
the same for a deterministic scale free network as well as
for a random scale free network. This implies a clear
advantage for an analytical treatment in a similar
way as was done for the Sznajd model~\cite{Gonzalez2006b}.

\section*{Acknowledgments}
This work is supported by the Max-Planck Price and the German agency DAAD as well as by the
Brasilian agencies CNPq, CAPES and FUNCAP.

\bibliographystyle{revtex}
\bibliography{penna}
\end{document}